\pdfoutput=1
\documentclass[aps,prc,reprint,preprintnumbers,superscriptaddress,floatfix,nofootinbib]{revtex4-2}
\usepackage{multirow}
\usepackage{graphicx}
\usepackage{color}
\usepackage{amsmath}
\usepackage{amssymb}
\usepackage[colorlinks=true]{hyperref}
\usepackage{tikz}
\usepackage{etoolbox}
\usepackage{bm}
\usepackage[utf8]{inputenc}
\usepackage{verbatim}
\usepackage[bottom]{footmisc}
\usepackage{comment}
\usepackage{booktabs}
\usepackage{array}
\usepackage{combelow}

\begin{document}
\preprint{RIKEN-iTHEMS-Report-26}

\title{Recurrent Fourier Neural Operator for Event-by-Event Relativistic Hydrodynamics
}

\author{Xiangrun Bao}
\email{2201110096@stu.pku.edu.cn}
\affiliation{School of Physics, Peking University, Beijing 100871, China}

\author{Lingxiao Wang}
\email{lingxiao.wang@riken.jp}
\affiliation{RIKEN Center for Interdisciplinary Theoretical and Mathematical Sciences (iTHEMS), Wako, Saitama 351-0198, Japan}
\affiliation{Institute for Physics of Intelligence, Graduate School of Science, The University of Tokyo, Bunkyo-ku, Tokyo 113-0033, Japan}
\author{Huichao Song}
\email{huichaosong@pku.edu.cn}
\affiliation{School of Physics, Peking University, Beijing 100871, China}
\affiliation{Center for High Energy Physics, Peking University, Beijing 100871, China}
\date{\today}

\begin{abstract}
Relativistic hydrodynamics is a powerful tool to simulate the evolution of the quark--gluon plasma (QGP) in relativistic heavy-ion collisions,  but event-by-event simulations are computationally demanding, especially for Bayesian analyses. We develop a recurrent Fourier neural operator (rFNO) that emulates the evolution of $(2+1)$-dimensional relativistic hydrodynamics. Three networks are trained on the $32\times32$,
$64\times64$, and $128\times128$ lattices using profiles of the same \textsc{VISH2+1} hydrodynamic events up to $\tau-\tau_0=4.8~\mathrm{fm}/c$. The networks trained on the coarser lattices are then directly evaluated on the $128\times128$ lattice without retraining, while the $128\times128$ network provides the native-resolution reference.  A comparison with the \textsc{VISH2+1} results shows that the rFNO predictions across resolutions nicely capture the magnitude and structures of the temperature profiles, and describe the event-by-event eccentricities $\varepsilon_n$ ($n=2,3,4$), the momentum anisotropy $\varepsilon_p$, and their time evolution within the supervised proper-time interval. After the recurrent continuation to $\tau-\tau_0=6.4~\mathrm{fm}/c$, the relative errors of the predicted $\varepsilon_n$ and $\varepsilon_p$ remain within about $10\%$. These results demonstrate that the rFNO realizes zero-shot super-resolution predictions for relativistic hydrodynamics and can be recurrently continued beyond the supervised interval.
\end{abstract}

\maketitle

\section{Introduction}
\label{sec:intro}

Probing the properties of the quark--gluon plasma (QGP) is one of the essential goals
of ultrarelativistic heavy-ion collision (HIC) experiments at the Relativistic Heavy Ion Collider (RHIC) and the Large Hadron Collider (LHC)~\cite{STAR:2005gfr,PHENIX:2004vcz,BRAHMS:2004adc,PHOBOS:2004zne,ALICE:2022wpn,Gyulassy:2004zy,Muller:2012zq,Shuryak:2014zxa}. Relativistic hydrodynamics has successfully described the collective expansion of the QGP and, when combined with models for the initial conditions, hadronization, and hadronic afterburner, provided a quantitative description of a broad range of bulk soft observables~\cite{Kolb:2003dz,TEANEY_2010,Heinz:2013th,Hirano:2012kj, Florkowski:2017olj,Song:2017wtw,Heinz:2024jwu}. However, event-by-event hydrodynamic simulations are computationally demanding when used to constrain model parameters through Bayesian analyses of experimental data, since the hydrodynamic simulations must be repeated for numerous points in the parameter space~\cite{Bernhard:2016tnd,Bernhard:2019bmu,JETSCAPE:2020shq,JETSCAPE:2020mzn,Paquet2023EmulationReview,Nijs2020TrajectumBayes}. Gaussian process emulators are commonly used to interpolate final-state observables across parameter space~\cite{Petersen:2010zt,Paquet2023EmulationReview,Roch2024EmulatorPRC}. In this paper, we construct a neural operator to predict the evolving hydrodynamic profiles at successive proper times from the initial profiles.

Machine learning methods have been extensively explored as emulators for HIC simulations and as tools for analyzing high-dimensional data~\cite{Karniadakis2021,Boehnlein2022MLNuclear,He2023HENPML,Zhou2023QCDML}, and large language model agents have recently been used to run and analyze relativistic hydrodynamic simulations of HICs~\cite{Wang2026CLViscAgent}. Deep neural networks (DNNs) have been used to infer properties of QCD matter from hydrodynamic simulations and to map initial conditions to hydrodynamic profiles at later proper times~\cite{Pang2018EoSMeter,Huang2021DLHydro}.  Meanwhile, physics-informed neural networks (PINNs) have also been applied to relativistic hydrodynamics~\cite{FerrerSanchez2024GAPINN,ChomaliCastro2026BDNK,Matsuda:2025whj}. However, DNNs learn mappings between finite-dimensional vector spaces, rendering their predictions fundamentally grid-dependent, whereas PINNs are mesh-free but in general must be retrained for each new initial condition.

Operator learning methods overcome these limitations by approximating mappings between infinite-dimensional function spaces, enabling discretization-invariant representations~\cite{Li2021FNO,Kovachki2023NO,Lu2021DeepONet,Azizzadenesheli2024NORP}. In particular, the Fourier neural operator (FNO) parameterizes integral kernel operators in Fourier space, efficiently capturing global spatial correlations~\cite{Li2021FNO,Kovachki2023NO}. The first application of FNOs to QGP hydrodynamics showed that a model trained on a $60\times60$ lattice can be applied to a $150\times150$ lattice with comparable accuracy~\cite{StewartPutschke2025FNOQGP}. However, this model maps the initial profile directly to the evolution over a fixed time window, while the recurrent propagation of intermediate hydrodynamic profiles and its transfer across resolutions remain unexplored.

In this paper, we develop a recurrent FNO (rFNO), a finite-horizon neural operator that propagates $(2+1)$-dimensional relativistic hydrodynamic profiles from one time interval to the next, and investigate its generalization in both spatial resolution and evolution time. Three rFNOs, referred to as the $32\times32$, $64\times64$, and $128\times128$ rFNOs, are trained on the corresponding lattices using the same event-by-event hydrodynamic profiles. The $32\times32$ and $64\times64$ rFNOs are then evaluated on the $128\times128$ lattice without retraining, with the $128\times128$ rFNO as the native-resolution reference. Comparisons with the hydrodynamic ground truth show that the rFNO achieves zero-shot super-resolution and can be recurrently continued beyond the supervised time interval, which could help accelerate realistic event-by-event hydrodynamic simulations.

\section{Hydrodynamics and rFNOs}
\label{sec:method}

\subsection{VISH2+1 hydrodynamics}
\label{subsec:data}
In this paper, we use the $(2+1)$-dimensional hydrodynamic code \textsc{VISH2+1}~\cite{Song2008} to simulate the event-by-event evolution of the QGP fireball, which generates the training and test profiles for the neural network. For simplicity, we run its ideal hydrodynamic mode with zero net baryon density, which solves the conservation equations of the energy--momentum tensor,
\begin{equation}
    \partial_\mu T^{\mu\nu}=0,
    \qquad
    T^{\mu\nu}=(e+P)u^\mu u^\nu-Pg^{\mu\nu},
\label{eq:conservation}
\end{equation}
where $e$ is the local energy density, $P$ is the local pressure, and $u^\mu$ is the local four-velocity. The equation of state is set to the conformal one, $P=e/3$. In general, it is convenient to work in Milne coordinates $x^\mu=(\tau,x,y,\eta_s)$, with the proper time $\tau=\sqrt{t^2-z^2}$ and the space--time rapidity $\eta_s=\frac{1}{2}\ln[(t+z)/(t-z)]$. With the assumption of longitudinal boost invariance $v_z = \frac{z}{t}$~\cite{Bjorken1983}, the hydrodynamic profiles are independent of $\eta_s$. VISH2+1 then simulates the evolution of $T^{\tau\tau}(\tau,x,y)$, $T^{\tau x}(\tau,x,y)$, and $T^{\tau y}(\tau,x,y)$, from which $e(\tau,x,y)$, $P(\tau,x,y)$, $u^\mu(\tau,x,y)$, and the local temperature $T(\tau,x,y)$ are reconstructed with the input equation of state for each time step.

The initial profiles are generated by the {T\raisebox{-0.5ex}{R}ENTo} model~\cite{Moreland2015} for Au+Au collisions at $\sqrt{s_{NN}}=200$~GeV in the $0$--$10\%$ centrality class, with the parameters tuned close to the ones extracted from the Bayesian analyses for Pb+Pb collisions at the LHC~\cite{Bernhard:2016tnd,Bernhard:2019bmu}. The pre-equilibrium evolution is described by free streaming, which generates nonzero transverse flow~\cite{Liu2015FreeStreaming}. The subsequent collective expansion of the QGP is described by \textsc{VISH2+1}~\cite{Song2008} from $\tau_0=0.6~\mathrm{fm}/c$ until $\tau-\tau_0=8~\mathrm{fm}/c$. The simulations are performed on a \(\Delta x=\Delta y=0.2~\mathrm{fm}\) lattice, with the time step \(\Delta\tau=0.08~\mathrm{fm}/c\).

For the neural network, the transverse profiles of the energy--momentum tensor and local
temperature at each proper time serve as both the inputs and the prediction targets, which
are treated as images with four channels
\begin{equation}
    \mathcal{T}(\tau,x,y)=\left(T^{\tau\tau},\,T^{\tau x},\,T^{\tau y},\,T\right).
\label{eq:channels}
\end{equation}
To study the spatial-resolution transfer ability of the network, we construct three sets of training and test profiles from the same hydrodynamic evolution profiles. The profiles on the $128\times128$ lattice are taken from the original hydrodynamic simulations, and the $64\times64$ and $32\times32$ profiles are obtained by an average over $2\times2$ and $4\times4$ neighboring cells, which give the lattice spacings of $0.2$, $0.4$, and $0.8~\mathrm{fm}$, respectively, as summarized in Table~\ref{tab:resolution_setup}.  In practice, we run 6000 event-by-event hydrodynamic simulations and divide them into 4500 training events, 500 validation events, and 1000 test events.

To evaluate the performance of the neural network, we also calculate the spatial eccentricities $\varepsilon_n$ ($n=2,3,4$) and the momentum anisotropy $\varepsilon_p$ from the hydrodynamic profiles~\cite{Kolb:2003dz,Heinz:2013th}. The spatial eccentricities
characterize the deformation and inhomogeneity of the fireball and are defined as
\begin{equation}
    \varepsilon_n=\frac{\left|\int r\,dr\,d\phi\, r^n e(r,\phi)e^{in\phi}
    \right|}{\int r\,dr\,d\phi\, r^n e(r,\phi)},\qquad n=2,3,4,
\label{eq:eccentricity}
\end{equation}
where $e(r,\phi)$ is the energy density in the transverse plane and $\phi$ is the azimuthal angle.

The momentum anisotropy characterizes the anisotropy of the collective flow and is defined as
\begin{equation}
    \varepsilon_p=\frac{{\int dx dy\left(T^{xx}-T^{yy}\right)}}
    {\int dx dy \left(T^{xx}+T^{yy}\right)}, \qquad \qquad \qquad \qquad
\label{eq:momentum_anisotropy}
\end{equation}
Here, $T^{xx}(\tau,x,y)$ and $T^{yy}(\tau,x,y)$ are taken from the hydrodynamic simulation, which are related to both the energy density and fluid velocity profiles.

\begin{figure*}[htbp!]
    \centering
    \includegraphics[width=0.8\textwidth]{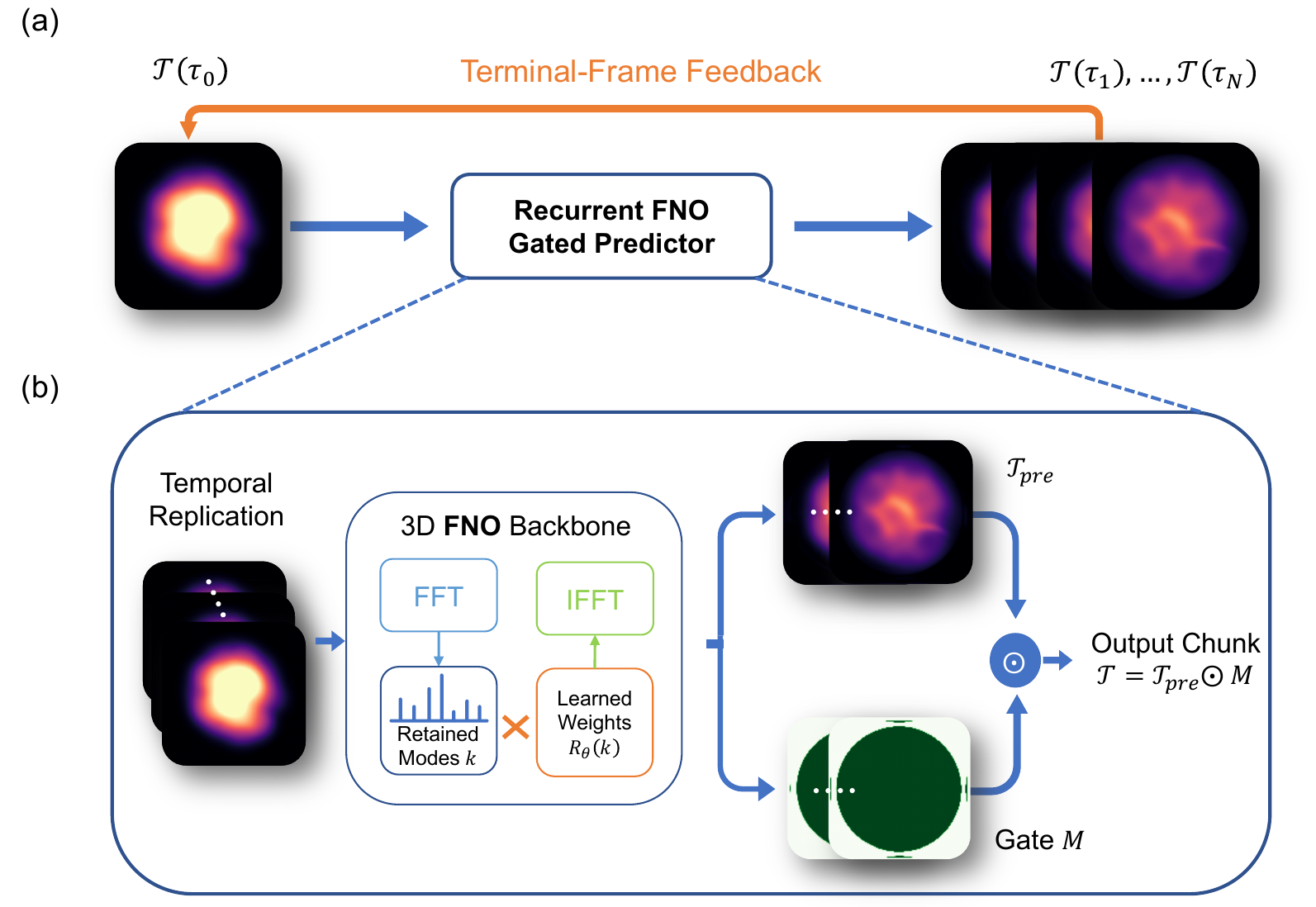}
    \caption{
    rFNO for the proper-time evolution of the hydrodynamic profiles. Panel (a) illustrates the recurrent procedure: starting from $\mathcal{T}(\tau_0)$, the network predicts the profiles over a finite proper-time interval and uses the final prediction to initialize the next interval. Panel (b) shows the prediction module for one interval. The input profiles are replicated along the temporal direction and processed by a three-dimensional FNO acting on $(\tau,x,y)$; a learned soft gate suppresses small-amplitude predictions in the dilute exterior region. The same procedure is used for the models trained on the $32\times32$, $64\times64$, and $128\times128$ transverse lattices.
    }
    \label{fig:overview}
\end{figure*}

\begin{figure*}[htbp!]
    \centering
\includegraphics[width=0.9\textwidth]{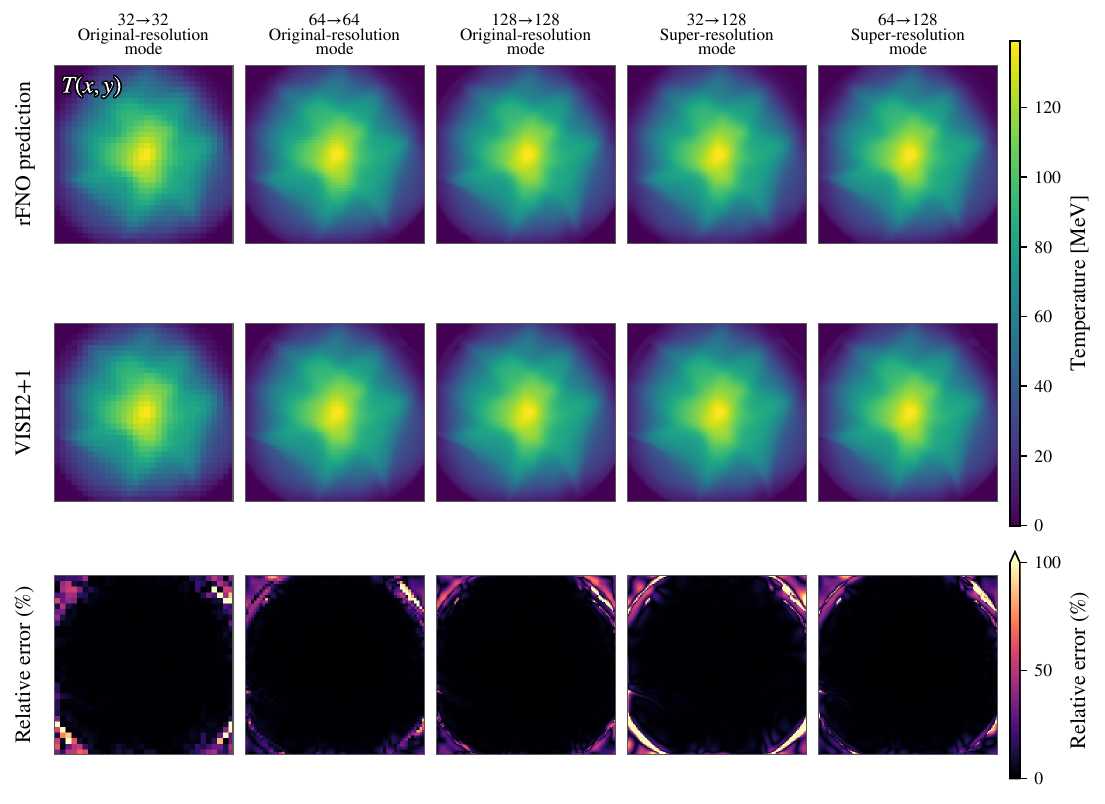}
    \caption{Temperature profiles at $\tau-\tau_0=4.8$ fm/$c$ for one selected event. From top to bottom, the rows show the rFNO prediction, the \textsc{VISH2+1} reference, and their relative error. The first three columns show the predictions of the $32\times32$, $64\times64$, and $128\times128$ rFNOs on their native lattices, and the fourth and fifth columns show the super-resolution predictions of the $32\times32$ and $64\times64$ rFNOs on the $128\times128$ lattice, obtained without retraining from the $128\times128$ initial profiles.   }
    \label{fig:spatial_transfer}
\end{figure*}

\begin{table}[htbp]
\centering
\small
\caption{rFNO lattice spacing, block-averaging factor, retained Fourier modes (temporal, two transverse directions), and evaluation lattices for the three trained models.}
\label{tab:resolution_setup}
\resizebox{\columnwidth}{!}{%
\begin{tabular}{ccccc}
\hline
Lattice & Spacing & Block & Modes & Evaluation lattices \\
\hline
$32\times32$ & $0.8$ fm & $4\times4$ & $(6,16,16)$ & $32^2$, $128^2$ \\
$64\times64$ & $0.4$ fm & $2\times2$ & $(6,32,32)$ & $64^2$, $128^2$ \\
$128\times128$ & $0.2$ fm & none & $(6,48,48)$ & $128^2$ \\
\hline
\end{tabular}
}
\end{table}

\subsection{rFNO and spatial-resolution transfer}
\label{subsec:backbone}
The network maps the hydrodynamic state at one proper time to a sequence of transverse profiles over a finite interval (see Fig.~\ref{fig:overview}). The input state is replicated along the temporal direction and processed by a three-dimensional FNO acting on $(\tau,x,y)$ with four Fourier layers and hidden width 64~\cite{Li2021FNO,Kovachki2023NO}. Normalized grid coordinates $(\tau,x,y)$ are included as additional input channels. The transverse Fourier transforms are performed on the grid without additional spatial padding. The final predicted time slice is used as the input for the next recurrent step. Table~\ref{tab:resolution_setup} summarizes the lattice and spectral settings. The shortest transverse wavelengths retained in the spectral layers are approximately $3.2$, $1.6$, and $1.1~\mathrm{fm}$ for the models trained on the $32\times32$, $64\times64$, and $128\times128$ lattices, respectively.  A learned output gate suppresses small-amplitude predictions in the dilute exterior region. The gating head uses a two-dimensional FNO with two layers acting on each time slice and produces a continuous sigmoid mask for every predicted variable, proper time, and transverse position. During cross-resolution evaluation, both the Fourier weights and the gating parameters remain unchanged. The final prediction is written as
\begin{equation}
    \widehat{\mathcal{T}}
    =
    \mathcal{M}\odot \widehat{\mathcal{T}}_{\rm pre},
\end{equation}
where $\widehat{\mathcal{T}}_{\rm pre}$ denotes the rFNO prediction before gating, $\mathcal{M}\in[0,1]$ the learned soft mask, and $\odot$ element-wise multiplication.

All models used in the spatial-resolution analysis employ the same temporal target construction and recurrent training protocol. We write $\mathcal{T}_j=\mathcal{T}(\tau_0+j\Delta\tau_{\mathrm{net}})$, where $\Delta\tau_{\mathrm{net}}=0.2~\mathrm{fm}/c$ is the spacing of the network time slices; each interval contains $N_t=8$ future profiles and spans $1.6~\mathrm{fm}/c$. The hydrodynamic profiles saved every $0.08~\mathrm{fm}/c$ are linearly interpolated in proper time onto these slices. Temporal reflection extends a sequence of $N_t+1$ profiles including the input profile according to
\begin{equation}
\left(\mathcal{T}_0,\ldots,\mathcal{T}_{N_t}\right)
\longrightarrow
\left(\mathcal{T}_0,\ldots,\mathcal{T}_{N_t},
\mathcal{T}_{N_t-1},\ldots,\mathcal{T}_0\right).
\label{eq:temporal_reflection}
\end{equation}
This construction reduces the mismatch at the temporal boundaries encountered in Fourier processing over a finite prediction interval. The auxiliary reflected portion is removed before the physical prediction sequence is evaluated.

For the native-resolution tests, each model is evaluated on the transverse lattice used during training. For the cross-resolution tests, the input profiles are represented directly on a finer lattice while all learned model parameters are kept unchanged. We consider the $32\times32\to128\times128$ and $64\times64\to128\times128$ transfers. The predictions are compared with the hydrodynamic reference represented on the corresponding evaluation lattice. No fine-tuning or model selection on the test set is performed for these cross-resolution evaluations.

\subsection{Progressive recurrent training}
\label{subsec:training}

During the recurrent training stages, the network receives its own predictions as inputs to account for errors accumulated during recurrent evolution. We progressively increase the training horizon, following rollout-aware training strategies for neural PDE models~\cite{Brandstetter2022MPNPDE,Liu2023MSMPPDE,Poli2023AutoregressivePDE}.

Denoting the three supervised temporal intervals by $I_1$, $I_2$, and $I_3$, with starting proper times $\tau_0$, $\tau_1$, and $\tau_2$, respectively, the training curriculum is
\begin{equation}
\begin{aligned}
\text{Stage I:}\;&\quad
\tau_0\!\to I_1,\quad
\tau_1\!\to I_2,\quad
\tau_2\!\to I_3,\\
\text{Stage II:}\;&\quad
\tau_0\!\to I_1\!\to I_2,\quad
\tau_1\!\to I_2\!\to I_3,\\
\text{Stage III:}\;&\quad
\tau_0\!\to I_1\!\to I_2\!\to I_3.
\end{aligned}
\label{eq:progressive_training}
\end{equation}
 The three stages are scheduled for 64, 64, and 128 epochs, respectively, giving 256 epochs in total. In Stages II and III, the final prediction of one interval initializes the next without teacher forcing, with shared network parameters. All three models use this curriculum.

The loss is the element-wise mean squared error over aligned predictions and targets, with the prediction error in each channel normalized by the corresponding root-mean-square (RMS) value calculated from the training set, plus an auxiliary binary cross-entropy (BCE) loss on the gate mask weighted by $\alpha=0.01$. The gate is optimized jointly with the predictor, using binary target masks defined as $m_{\mathrm{target}}=\mathbf{1}(|u_{\mathrm{target}}|>10^{-6})$, where $u_{\mathrm{target}}$ is the target already normalized by the RMS of each channel over the training set. We use Adam with initial learning rate $10^{-3}$ and batch size 2, and select the single checkpoint with the lowest global relative $L_2$ error on the validation set for evaluation on the independent test set.

\begin{figure*}[t]
    \centering
    \includegraphics[width=\textwidth]{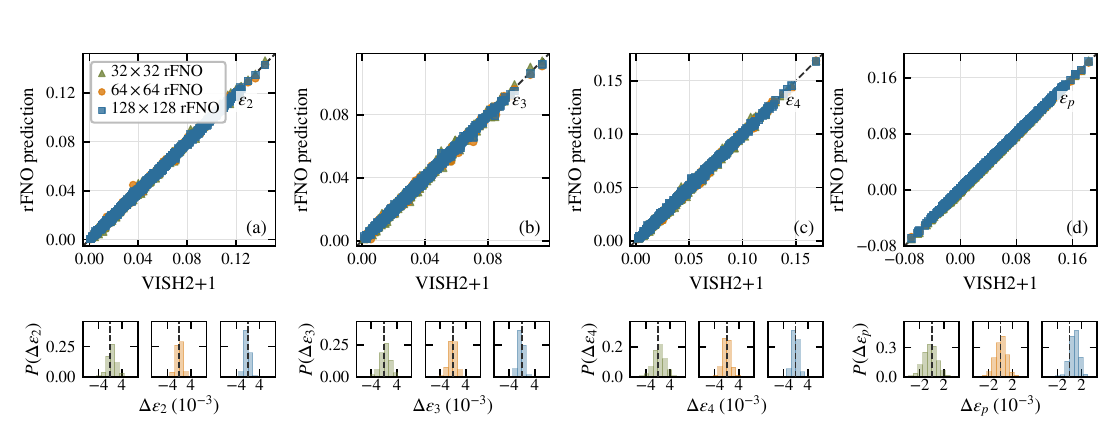}
    \caption{Event-by-event comparison of the spatial eccentricities $\varepsilon_2$, $\varepsilon_3$, and $\varepsilon_4$ and the momentum anisotropy $\varepsilon_p$ at $\tau-\tau_0=4.8~\mathrm{fm}/c$ between the rFNO predictions and the \textsc{VISH2+1} results (upper panels), together with the distributions of their differences $\Delta\varepsilon=\varepsilon_{\rm rFNO}-\varepsilon_{\rm VISH2+1}$ (lower panels). The $32\times32$ and $64\times64$ rFNOs are evaluated on the $128\times128$ lattice, and the $128\times128$ rFNO is evaluated on its native lattice.}
    \label{fig:integrated_observables}
\end{figure*}

\section{Results and Discussion}
As described in Sec.~\ref{sec:method}, we trained three rFNOs on the $32\times32$, $64\times64$, and $128\times128$ lattices, using the $\mathcal{T}(\tau,x,y)=\left(T^{\tau\tau},\,T^{\tau x},\,T^{\tau y},\,T\right)$ profiles  generated from event-by-event \textsc{VISH2+1} simulations. In this section, we compare the rFNO predictions with the \textsc{VISH2+1} results for the temperature profiles, the event-by-event eccentricities and momentum anisotropy, with a focus on the super-resolution predictions of the $32\times32$ and $64\times64$ rFNOs on the $128\times128$ lattice and on the predictions beyond the supervised interval.

Figure~\ref{fig:spatial_transfer} shows the temperature profiles from the rFNO predictions and from \textsc{VISH2+1} simulations, for a representative test event whose prediction error is close to the median of the test sample. The bottom panels show the relative error, which is defined for each grid point as $|T_{\rm rFNO}-T_{\rm VISH2+1}|/T_{\rm VISH2+1}$. Here we test these predictions at $\tau-\tau_0=4.8~\mathrm{fm}/c$, which is the end of the supervised interval. The first three columns of Fig.~\ref{fig:spatial_transfer} show the results of the $32\times32$, $64\times64$, and $128\times128$ rFNOs on their native lattices, while the fourth and fifth columns show the super-resolution predictions of the $32\times32$ and $64\times64$ rFNOs on the $128\times128$ lattice. On their native lattices, all three models reproduce the magnitude and detailed structure of the temperature profiles, with very small relative errors within the interior region. The outside region shows large relative errors, but the temperature there has already dropped below 5 MeV, far below the freeze-out temperatures of about 100--150 MeV used for soft observables in realistic heavy-ion simulations.  When the $32\times32$ and $64\times64$ rFNOs are applied to the $128\times128$ initial profiles for super-resolution predictions, the predicted temperature profiles still capture the overall structure of the \textsc{VISH2+1} profiles, with the finer details resolved, even though these rFNOs have never seen the $128\times128$ profiles during training. The hot interior remains well described and the residual differences still concentrate on the edge of the fireball, where the temperature drops below $5$--$10$ MeV. Such differences are more pronounced for the $32\times32$ rFNO than for the $64\times64$ one. This may be attributed to the larger number of retained Fourier modes of the $64\times64$ rFNO (see Table~\ref{tab:resolution_setup}) and its finer training lattice, which better resolve the temperature gradients. These results demonstrate that the rFNO trained on a coarse lattice can make super-resolution predictions.

\begin{figure*}[t]
    \centering
    \begin{minipage}[t]{0.3\textwidth}
        \centering
        \includegraphics[width=\linewidth]{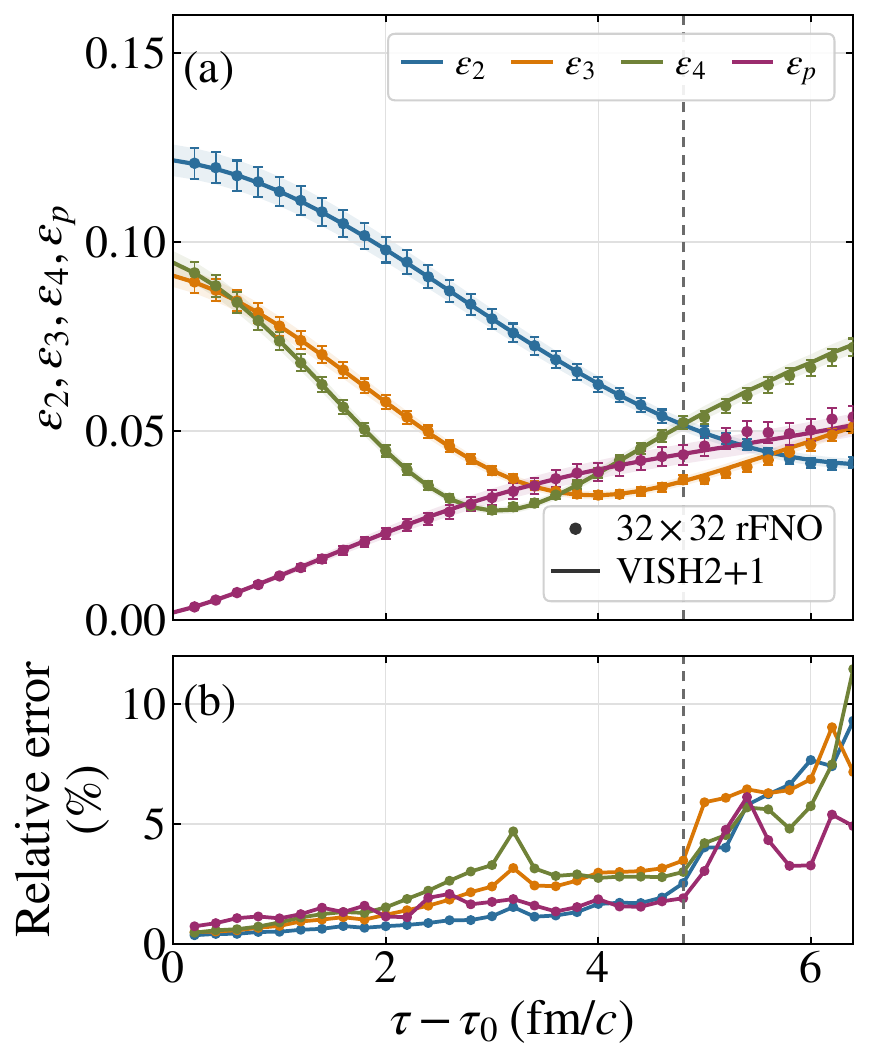}
    \end{minipage}
    \hfill
    \begin{minipage}[t]{0.3\textwidth}
        \centering
        \includegraphics[width=\linewidth]{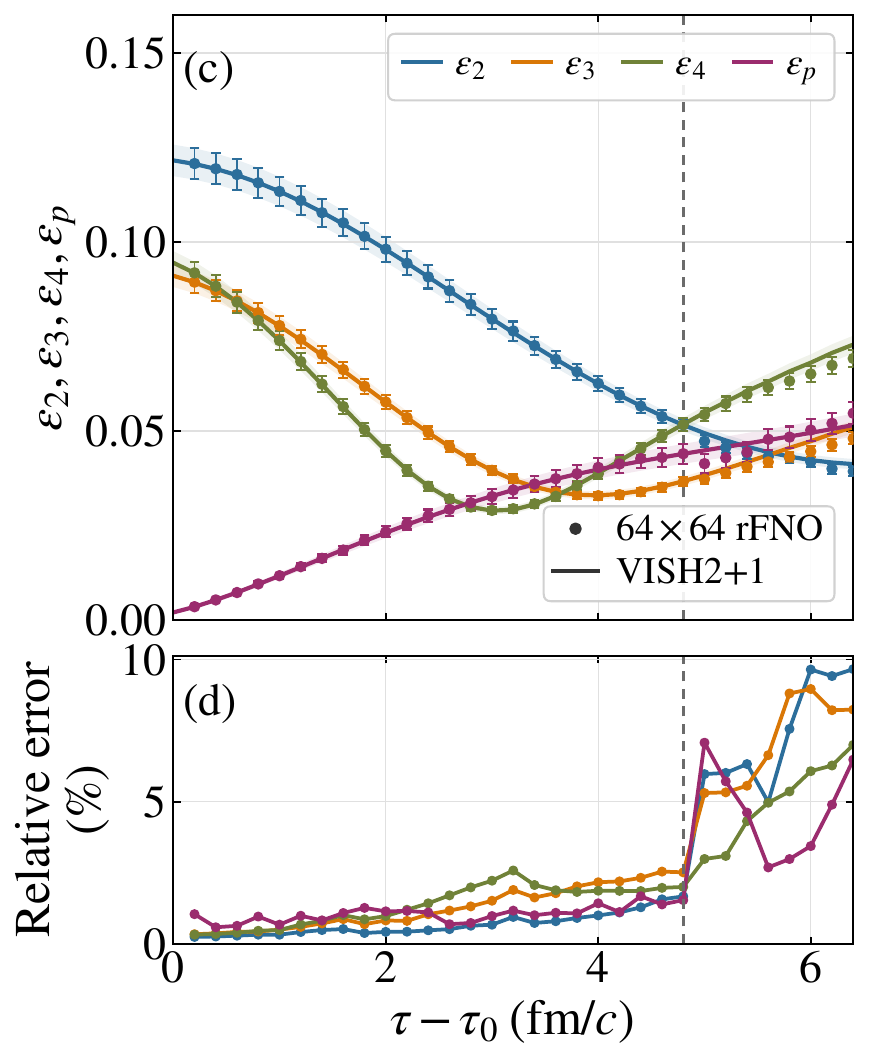}
    \end{minipage}
    \hfill
    \begin{minipage}[t]{0.3\textwidth}
        \centering
        \includegraphics[width=\linewidth]{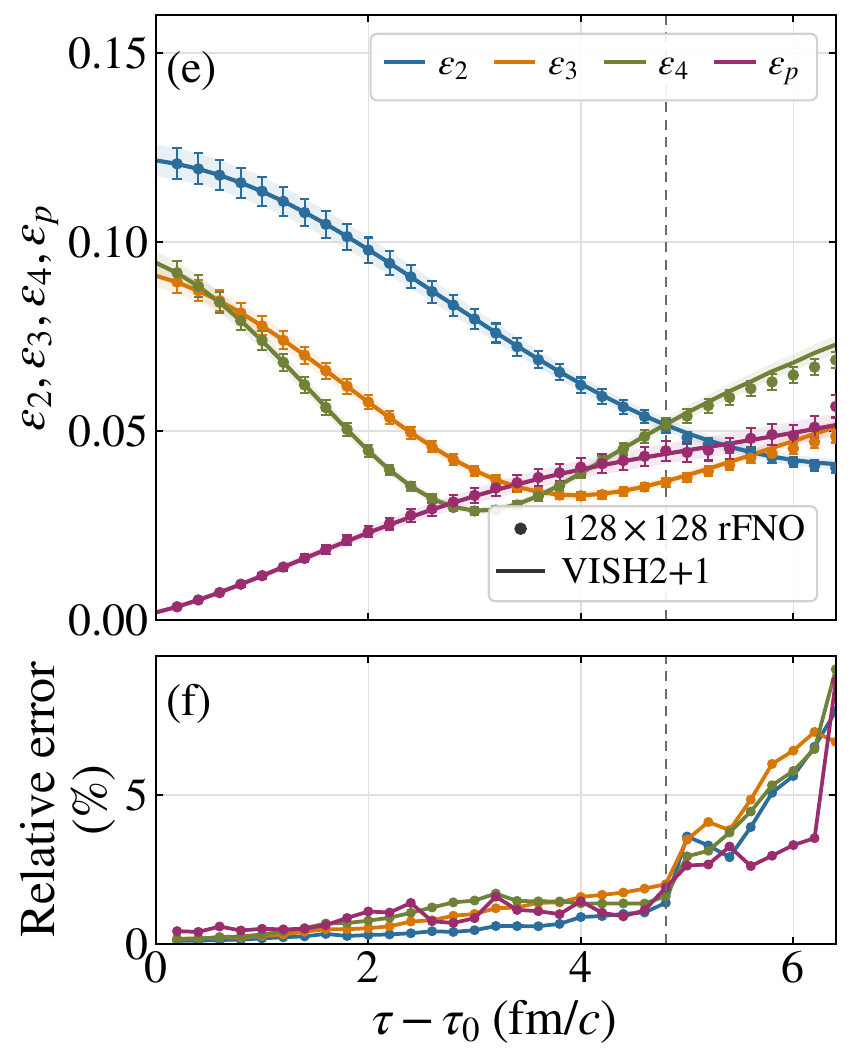}
    \end{minipage}
    \caption{Time evolution of $\langle\varepsilon_n\rangle$ ($n=2,3,4$) and $\langle\varepsilon_p\rangle$ averaged over 1000 events for the rFNO predictions and the \textsc{VISH2+1} ground truth. The left and middle panels show the super-resolution predictions of the $32\times32$ and $64\times64$ rFNOs on the $128\times128$ lattice, respectively, and the right panel shows the native-resolution predictions of the $128\times128$ rFNO. The vertical dashed lines mark the end of the supervised interval.}
    \label{fig:time_observables}
\end{figure*}

Figure~\ref{fig:integrated_observables} shows the event-by-event comparison of the spatial eccentricities $\varepsilon_n$ ($n=2,3,4$) and the momentum anisotropy $\varepsilon_p$ between the rFNO predictions (vertical axes) and the \textsc{VISH2+1} results (horizontal axes) at the end of the supervised interval $\tau-\tau_0=4.8~\mathrm{fm}/c$. Here $\varepsilon_n$ and $\varepsilon_p$ have been defined in Sec.~\ref{subsec:data}, which characterize the spatial deformation and inhomogeneity of the fireball and the momentum anisotropy of its collective expansion, respectively. For all three rFNO models evaluated on the $128\times128$ lattice (natively for the $128\times128$ rFNO), the points fall closely on the diagonal over the whole range covered by these test events with different fluctuating geometries. Notably, the super-resolution predictions of the $64\times64$ and $32\times32$ rFNOs on the $128\times128$ lattice also reproduce $\varepsilon_n$ and $\varepsilon_p$, which are defined through integrals over the transverse plane for a single event. This also indicates that the residual differences in Fig.~\ref{fig:spatial_transfer}, which concentrate along the edge of the fireball, carry little weight in $\varepsilon_n$ and $\varepsilon_p$, because the energy density there is small ($e\propto T^4$). The lower panels of Fig.~\ref{fig:integrated_observables} further show the distributions of the differences $\Delta\varepsilon_n$ and $\Delta\varepsilon_p$ between the rFNO predictions and the \textsc{VISH2+1} results. Note that the distributions become narrower from the $32\times32$ to the $64\times64$ rFNO and are the narrowest for the $128\times128$ rFNO, which shows that the super-resolution predictions remain accurate for these integrated quantities, while the native $128\times128$ predictions remain the most accurate.

Figure~\ref{fig:time_observables} plots the time evolution of the average eccentricities $\langle\varepsilon_n\rangle$ with $n=2,3,4$ and the average momentum anisotropy $\langle\varepsilon_p\rangle$ averaged over 1000 events from the independent test set for the rFNO predictions and for the \textsc{VISH2+1} ground truth. The bottom panels plot the relative error, which is defined as $\sum|\varepsilon_{\rm rFNO}-\varepsilon_{\rm VISH2+1}|/\sum\varepsilon_{\rm VISH2+1}$, with the sums running over the events. The vertical dashed lines mark the end of the supervised interval at $\tau-\tau_0=4.8~\mathrm{fm}/c$, beyond which the predictions are obtained by recurrent continuation without additional training.
Here, the $32\times32$ and $64\times64$ rFNOs are evaluated directly on the $128\times128$ lattice, and the $128\times128$ rFNO is evaluated on its native lattice.  In all three cases, the recurrent prediction is continued beyond the supervised interval to $\tau-\tau_0=6.4~\mathrm{fm}/c$ without any additional training, which tests the ability of the rFNO to extrapolate beyond the trained time window. For the \textsc{VISH2+1} results, $\langle\varepsilon_2\rangle$ decreases monotonically while $\langle\varepsilon_p\rangle$ increases monotonically, reflecting the conversion of the initial spatial anisotropy into momentum anisotropy by anisotropic pressure gradients~\cite{Kolb:2003dz}; $\langle\varepsilon_3\rangle$ and $\langle\varepsilon_4\rangle$ pass through minima within the supervised interval and then rise again. Within the supervised interval, all three rFNO models reproduce these trends, and the relative errors generally grow to a few percent due to the accumulation of errors along the recurrent evolution. Beyond the supervised interval $\tau-\tau_0=4.8~\mathrm{fm}/c$, the predictions of all three rFNOs still follow the \textsc{VISH2+1} trends, although the errors increase more rapidly at late times. Among the three models, the $128\times128$ rFNO generally gives the smallest late-time errors for $\varepsilon_2$ and $\varepsilon_3$. These results demonstrate that the rFNO can continue the hydrodynamic evolution beyond the supervised interval with the relative errors reaching about $10\%$ at the end of the evolution.

To examine energy--momentum conservation in the predicted evolution, we calculate
\begin{equation}
\begin{aligned}
E(\tau)&=\tau\int_\Omega d^2x_\perp\,T^{\tau\tau},\\
P_i(\tau)&=\tau\int_\Omega d^2x_\perp\,T^{\tau i},
\qquad i=x,y,\\
W_L(\tau)&=\int_{\tau_0}^{\tau}d\tau'
\int_\Omega d^2x_\perp\,P(\tau',\mathbf{x}_\perp),\\
W_\perp(\tau)&=\int_{\tau_0}^{\tau}d\tau'\,\tau'
\oint_{\partial\Omega}d\ell
\sum_{i=x,y}T^{\tau i}(\tau',\mathbf{x}_\perp)n_i.
\end{aligned}
\label{eq:integrated_energy_work}
\end{equation}
Here $\Omega$ is the fixed transverse integration region, and $n_i$ are the components of the outward unit normal to its boundary $\partial\Omega$. The quantities $W_L$ and $W_\perp$ represent the longitudinal work and the net energy carried out through the transverse boundary, respectively. With $E_0=E(\tau_0)$, we measure the energy deviation $\delta_E$ and the transverse-momentum deviation $\delta_P$ using
\begin{equation}
\begin{aligned}
\delta_E(\tau)&=\left\langle
\frac{\left|E(\tau)+W_L(\tau)+W_\perp(\tau)-E_0\right|}{E_0}
\right\rangle,\\
\delta_P(\tau)
&=\sum\limits_{i=x,y}
\left\langle
\frac{\left|P_i(\tau)-P_i(\tau_0)\right|}{E_0}
\right\rangle.
\end{aligned}
\label{eq:global_balance_diagnostics}
\end{equation}

We use 1000 test events and evaluate all models on the $128\times128$ lattice (natively for the $128\times128$ rFNO). Table~\ref{tab:energy_momentum} summarizes the results at the end of the supervised interval, $\tau-\tau_0=4.8~\mathrm{fm}/c$, and after recurrent continuation to $\tau-\tau_0=6.4~\mathrm{fm}/c$. Although energy--momentum conservation is not explicitly imposed during training, the rFNO values of $\delta_E$ and $\delta_P$ are of the same order as those of \textsc{VISH2+1}, whose nonzero $\delta_E$ mainly reflects the different time discretizations used for the longitudinal pressure source and the evaluation of $W_L$, since $E+W_L+W_\perp=E_0$ holds exactly in ideal hydrodynamics. The smaller $\delta_E$ of the $128\times128$ rFNO at $\tau-\tau_0=6.4~\mathrm{fm}/c$ results from a partial cancellation between its prediction error and this discretization error, rather than from better energy conservation.

\begin{table}[tb]
\centering
\caption{Energy and transverse-momentum deviations. The lattice sizes indicate the training grids.}
\label{tab:energy_momentum}
\small
\setlength{\tabcolsep}{4pt}
\begin{tabular}{lcccc}
\hline
$\tau-\tau_0$ & \multicolumn{2}{c}{$4.8~\mathrm{fm}/c$} & \multicolumn{2}{c}{$6.4~\mathrm{fm}/c$} \\
Model & $\delta_E$ (\%) & $\delta_P$ (\%) & $\delta_E$ (\%) & $\delta_P$ (\%) \\
\hline
\textsc{VISH2+1} & 1.71 & 0.042 & 1.77 & 0.235 \\
rFNO ($32\times32$) & 1.37 & 0.077 & 2.02 & 0.502 \\
rFNO ($64\times64$) & 1.76 & 0.076 & 2.12 & 0.203 \\
rFNO ($128\times128$) & 1.68 & 0.059 & 0.23 & 0.225 \\
\hline
\end{tabular}
\end{table}

The rFNO predictions are also much faster than the \textsc{VISH2+1} simulations. The prediction times per event for the $32\times32$, $64\times64$, and $128\times128$ rFNOs are approximately $50$, $55$, and $64$~ms, respectively, when evaluated on the $128\times128$ lattice over $\tau-\tau_0=4.8~\mathrm{fm}/c$ using an NVIDIA GeForce RTX 4090 D GPU with batch size one. For comparison, generating a hydrodynamic event with \textsc{VISH2+1} over the same proper-time range takes approximately $25$~s on average using a single thread on an Intel Xeon Gold 6338 CPU. GPU parallelization of hydrodynamic evolution in CLVisc achieves a speedup of approximately $60$ relative to execution on a single CPU core~\cite{Pang2018CLVisc}. Assuming the same acceleration for \textsc{VISH2+1}, its estimated runtime would be approximately $0.42$~s per event, leaving the rFNO predictions on the $128\times128$ lattice about $6$--$8$ times faster. Implementing the $64\times64$ rFNO on the $128\times128$ lattice is expected to further  reduce the prediction time by a factor of 4.

\section{Summary and Conclusion}
\label{sec:conclusion}

In this work, we developed a recurrent Fourier neural operator (rFNO) that emulates the evolution of $(2+1)$-dimensional relativistic hydrodynamics, which is trained with the profiles generated from \textsc{VISH2+1} hydrodynamics with {T\raisebox{-0.5ex}{R}ENTo} initial conditions. Using the same hydrodynamic profiles up to $\tau-\tau_0=4.8~\mathrm{fm}/c$, three networks are trained on the $32\times32$, $64\times64$, and $128\times128$ lattices, and the two trained on the coarser lattices are then directly evaluated on the $128\times128$ lattice without retraining. A comparison with the \textsc{VISH2+1} results shows that the rFNO predictions nicely capture the magnitude and detailed structures of the temperature profiles, on both the native lattices and the finer $128\times128$ lattice. Such super-resolution predictions further nicely reproduce the event-by-event eccentricities $\varepsilon_n$ ($n=2,3,4$) and the momentum anisotropy $\varepsilon_p$ at the end of the supervised interval, $\tau-\tau_0=4.8~\mathrm{fm}/c$, as well as the time evolution of their event averages within this interval. Through recurrent continuation, the predictions roughly follow the \textsc{VISH2+1} evolution beyond the supervised interval, with the relative errors of $\varepsilon_n$ and $\varepsilon_p$ reaching about $10\%$ at $\tau-\tau_0=6.4~\mathrm{fm}/c$. We also found that the predicted evolution approximately satisfies global energy--momentum conservation with deviations of $<3\%$, the same order as those from \textsc{VISH2+1}. These results demonstrate that the rFNO
could make reliable super-resolution predictions and could be further recurrently continued beyond the supervised interval, which shows its potential to substantially accelerate realistic event-by-event hydrodynamic simulations. The time-resolved profiles predicted by the rFNO are complementary to recent generative diffusion and flow-matching models that emulate the final-state hadron spectra of hybrid hydrodynamic simulations end to end~\cite{Sun2024DiffHIC,Wu2026FlowMatchingJet}, and combining the two approaches is a promising direction for future work.

\begin{acknowledgments}
We thank Shigehiro Nagataki, Xingyu Guo, Shuzhe Shi, and Zhenyu Zhu for helpful discussions.
We thank the DEEP-IN working group at RIKEN-iTHEMS for support in the preparation of this paper.
XB and HS are supported by the NSFC under grant No.12575138, and No.12247107. LW is supported by the JSPS KAKENHI Grant No. 25H01560, and JST-BOOST Grant No. JPMJBY24H9. This work was performed at the Tianhe Supercomputer Center in Tianjin (Tianhe New Generation Supercomputer), and the High-Performance Computing Platform of Peking University.
\end{acknowledgments}

\bibliography{apssamp}

\end{document}